
%
%
%
%

\magnification=1200
\hsize=15truecm
\vsize=23truecm
\baselineskip 20 truept
\voffset=-0.5 truecm
\parindent=1cm
\overfullrule=0pt

%
%
%
%
\catcode`@=11
%
%
\def\b@lank{ }

\newif\if@simboli
\newif\if@riferimenti
\newif\if@bozze

\newwrite\file@simboli
\def\simboli{
    \immediate\write16{ !!! Genera il file \jobname.SMB }
    \@simbolitrue\immediate\openout\file@simboli=\jobname.smb}

\newwrite\file@ausiliario
\def\riferimentifuturi{
    \immediate\write16{ !!! Genera il file \jobname.AUX }
    \@riferimentitrue\openin1 \jobname.aux
    \ifeof1\relax\else\closein1\relax\input\jobname.aux\fi
    \immediate\openout\file@ausiliario=\jobname.aux}

\def\bozze{\@bozzetrue\font\tt@bozze=cmtt8}

\newcount\eq@num\global\eq@num=0
\newcount\sect@num\global\sect@num=0

\newif\if@ndoppia
\def\numerazionedoppia{\@ndoppiatrue\gdef\la@sezionecorrente{\the\sect@num}}

\def\se@indefinito#1{\expandafter\ifx\csname#1\endcsname\relax}
\def\spo@glia#1>{} 

\newif\if@primasezione
\@primasezionetrue

\def\s@ection#1\par{\immediate
    \write16{#1}\if@primasezione\global\@primasezionefalse\else\goodbreak
    \vskip\spaziosoprasez\fi\noindent
    {\bf#1}\nobreak\vskip\spaziosottosez\nobreak\noindent}
%

\def\sezpreset#1{\global\sect@num=#1
    \immediate\write16{ !!! sez-preset = #1 }   }

\def\spaziosoprasez{25pt}

\def\spaziosottosez{15pt}

\def\sref#1{\se@indefinito{@s@#1}\immediate\write16{ ??? \string\sref{#1}
    non definita !!!}
    \expandafter\xdef\csname @s@#1\endcsname{??}\fi\csname @s@#1\endcsname}

\def\autosez#1#2\par{
    \global\advance\sect@num by 1\if@ndoppia\global\eq@num=0\fi
    \xdef\la@sezionecorrente{\the\sect@num}
    \def\usa@getta{1}\se@indefinito{@s@#1}\def\usa@getta{2}\fi
    \expandafter\ifx\csname @s@#1\endcsname\la@sezionecorrente\def
    \usa@getta{2}\fi
    \ifodd\usa@getta\immediate\write16
      { ??? possibili riferimenti errati a \string\sref{#1} !!!}\fi
    \expandafter\xdef\csname @s@#1\endcsname{\la@sezionecorrente}
    \immediate\write16{\la@sezionecorrente. #2}
    \if@simboli
      \immediate\write\file@simboli{ }\immediate\write\file@simboli{ }
      \immediate\write\file@simboli{  Sezione
                                  \la@sezionecorrente :   sref.   #1}
      \immediate\write\file@simboli{ } \fi
    \if@riferimenti
      \immediate\write\file@ausiliario{\string\expandafter\string\edef
      \string\csname\b@lank @s@#1\string\endcsname{\la@sezionecorrente}}\fi
    \goodbreak\vskip\spaziosoprasez
    \noindent\if@bozze\llap{\tt@bozze#1\ }\fi
      {\bf\the\sect@num.\quad #2}\par\nobreak\vskip\spaziosottosez
    \nobreak\noindent}

\def\semiautosez#1#2\par{
    \gdef\la@sezionecorrente{#1}\if@ndoppia\global\eq@num=0\fi
    \if@simboli
      \immediate\write\file@simboli{ }\immediate\write\file@simboli{ }
      \immediate\write\file@simboli{  Sezione ** : sref.
          \expandafter\spo@glia\meaning\la@sezionecorrente}
      \immediate\write\file@simboli{ }\fi
    \s@ection#2\par}


\def\eqpreset#1{\global\eq@num=#1
     \immediate\write16{ !!! eq-preset = #1 }     }

\def\eqref#1{\se@indefinito{@eq@#1}
    \immediate\write16{ ??? \string\eqref{#1} non definita !!!}
    \expandafter\xdef\csname @eq@#1\endcsname{??}
    \fi\csname @eq@#1\endcsname}

\def\eqlabel#1{\global\advance\eq@num by 1
    \if@ndoppia\xdef\il@numero{\la@sezionecorrente.\the\eq@num}
       \else\xdef\il@numero{\the\eq@num}\fi
    \def\usa@getta{1}\se@indefinito{@eq@#1}\def\usa@getta{2}\fi
    \expandafter\ifx\csname @eq@#1\endcsname\il@numero\def\usa@getta{2}\fi
    \ifodd\usa@getta\immediate\write16
       { ??? possibili riferimenti errati a \string\eqref{#1} !!!}\fi
    \expandafter\xdef\csname @eq@#1\endcsname{\il@numero}
    \if@ndoppia
       \def\usa@getta{\expandafter\spo@glia\meaning
       \la@sezionecorrente.\the\eq@num}
       \else\def\usa@getta{\the\eq@num}\fi
    \if@simboli
       \immediate\write\file@simboli{  Equazione
            \usa@getta :  eqref.   #1}\fi
    \if@riferimenti
       \immediate\write\file@ausiliario{\string\expandafter\string\edef
       \string\csname\b@lank @eq@#1\string\endcsname{\usa@getta}}\fi}

\def\autoreqno#1{\eqlabel{#1}\eqno(\csname @eq@#1\endcsname)
       \if@bozze\rlap{\tt@bozze\ #1}\fi}
\def\autoleqno#1{\eqlabel{#1}\leqno\if@bozze\llap{\tt@bozze#1\ }
       \fi(\csname @eq@#1\endcsname)}

\def\numeriadestra{\let\autoeqno=\autoreqno}
\def\numaeriasinistra{\let\autoeqno=\autoleqno}
\numeriadestra

\newcount\cit@num\global\cit@num=0

\newwrite\file@bibliografia
\newif\if@bibliografia
\@bibliografiafalse

\def\lp@cite{[}
\def\rp@cite{]}
\def\trap@cite#1{\lp@cite #1\rp@cite}
\def\lp@bibl{[}
\def\rp@bibl{]}
\def\trap@bibl#1{\lp@bibl #1\rp@bibl}

\def\refe@renza#1{\if@bibliografia\immediate        
    \write\file@bibliografia{
    \string\item{\trap@bibl{\cref{#1}}}\string
    \bibl@ref{#1}\string\bibl@skip}\fi}

\def\ref@ridefinita#1{\if@bibliografia\immediate\write\file@bibliografia{
    \string\item{?? \trap@bibl{\cref{#1}}} ??? tentativo di ridefinire la
      citazione #1 !!! \string\bibl@skip}\fi}

\def\bibl@ref#1{\se@indefinito{@ref@#1}\immediate
    \write16{ ??? biblitem #1 indefinito !!!}\expandafter\xdef
    \csname @ref@#1\endcsname{ ??}\fi\csname @ref@#1\endcsname}

\def\c@label#1{\global\advance\cit@num by 1\xdef            
   \la@citazione{\the\cit@num}\expandafter
   \xdef\csname @c@#1\endcsname{\la@citazione}}

\def\bibl@skip{\vskip 0truept}


\def\stileincite#1#2{\global\def\lp@cite{#1}\global
    \def\rp@cite{#2}}
\def\stileinbibl#1#2{\global\def\lp@bibl{#1}\global
    \def\rp@bibl{#2}}

\def\citpreset#1{\global\cit@num=#1
    \immediate\write16{ !!! cit-preset = #1 }    }

\def\autobibliografia{\global\@bibliografiatrue\immediate
    \write16{ !!! Genera il file \jobname.BIB}\immediate
    \openout\file@bibliografia=\jobname.bib}

\def\cref#1{\se@indefinito                  
   {@c@#1}\c@label{#1}\refe@renza{#1}\fi\csname @c@#1\endcsname}

\def\cite#1{\trap@cite{\cref{#1}}}                  
\def\ccite#1#2{\trap@cite{\cref{#1},\cref{#2}}}     
\def\ncite#1#2{\trap@cite{\cref{#1}--\cref{#2}}}    
\def\upcite#1{$^{\,\trap@cite{\cref{#1}}}$}               
\def\upccite#1#2{$^{\,\trap@cite{\cref{#1},\cref{#2}}}$}  
\def\upncite#1#2{$^{\,\trap@cite{\cref{#1}-\cref{#2}}}$}  

\def\clabel#1{\se@indefinito{@c@#1}\c@label           
    {#1}\refe@renza{#1}\else\c@label{#1}\ref@ridefinita{#1}\fi}

\def\biblskip#1{\def\bibl@skip{\vskip #1}}           

\def\insertbibliografia{\if@bibliografia             
    \immediate\write\file@bibliografia{ }
    \immediate\closeout\file@bibliografia
    \catcode`@=11\input\jobname.bib\catcode`@=12\fi}


\def\commento#1{\relax}
\def\biblitem#1#2\par{\expandafter\xdef\csname @ref@#1\endcsname{#2}}


\catcode`@=12


\autobibliografia
\numerazionedoppia
\def\au{\autoeqno}
\def\e#1{(\eqref{#1})}

\autosez{1} Introduction and motivations\par

Supergravity (SUGRA) theories arose as attempts towards a unification of the
fundamental interactions, including Quantum Gravity, and with this respect
their role has been confirmed with the advent of superstring theories
 and, more
speculatively, of the theory of supersymmetric extended objects, called
super $p$-branes. A super $p$-brane lives on a $(p+1)$-dimensional
world-sheet in
a $D$-dimensional target  super-space-time; the string has then to be
considered as a 1-brane $(p=1)$. The allowed values of $D$, for a given $p$,
are dictated by classical space-time supersymmetry \ccite{ACET}{DUFF}
and may be further restricted by consistency requirements at the quantum
level. At low energies these theories can be described by SUGRA
theories in $D$ space-time dimensions and these SUGRA theories describe also
the target space dynamics of the super $p$-brane $\sigma$-models. In the target
space  one can also have extended $N=2$ supersymmetry, see \cite{DULU2}, but
in this paper we concentrate on theories with simple $N=1$ space-time
supersymmetry.

One of the remarkable features which arose recently in the physics of extended
objects is the string ($p=1$, $D=10$) -- five-brane ($p=5$, $D=10$) duality
\ccite{STRO}{DLU}, meaning essentially that one theory can be regarded as a
soliton solution of the other. According to a strong version of the duality
conjecture \ccite{DUFF}{DDP} the two theories are equivalent in the sense that
they are just different mathematical descriptions of the same underlying
physics. The same should then also be true for the two corresponding $N=1$,
$D=10$ supergravity theories.

In this paper we present a unified formulation
of the two {\it pure} SUGRA theories which arise respectively as background
theories of the string and five-brane $\sigma$-models at the classical level.
The first SUGRA theory is usually described in terms of a closed
three-superform
$H_3$ \cite{NI} (corresponding to the string) and the second (dual) theory in
terms of a closed seven-superform $H_7$ \cite{GNISH} (corresponding to the
five-brane).

We discuss the issue of duality also in {\it non minimal}
$N=1$, $D=10$ SUGRA theories which take quantum corrections to
the heterotic {\it string} $\sigma$-model into account.
In this case, in particular,
the differential of $H_3$ is proportional to a second order polynomial
in the gauge and Lorentz curvatures, $dH_3=Tr(F^2)-tr(R^2)$, while
$H_7$ remains closed.
We leave the
discussion of the SUGRA theory where the differential of $H_7$ becomes
proportional
to a fourth order polynomial in the curvatures while $H_3$ remains closed,
which takes quantum corrections to the heterotic {\it five-brane}
$\sigma$-model into account, to a future publication \cite{CL1}.

 The {\it super\/}forms $H_3$ and $H_7$ are related in a way which
resembles much the duality relation between three and seven-forms in
{\it ordinary} ten-dimensional space and are therefore usually said to
be ``dual" to each other.
The two theories, which are known to be equivalent,
are most conveniently described in superspace. One has to choose
an appropriate set of constraints on curvatures and torsions and then to solve
the Bianchi identities. In the current treatments in the literature, according
to the set of constraints one uses, one has to impose the Bianchi identity
$dH_3=0$ to set the theory on shell
\ccite{NI}{ADR} while the identity $dH_7=0$
does not
contain any dynamical information and, in particular, does not set the theory
on shell \cite{DF}.

In the new formulation of $D=10$, $N=1$ SUGRA which we present here none of
these identities are imposed as starting points, they are rather both
consequences of the (simple) constraint we will impose on the
super-Riemann curvature and, moreover, in this formulation the fields
$H_3$ and $H_7$ are not introduced explicitly ``by hand'' at the beginning;
the (closed) forms $H_3$ and $H_7$ will arise naturally as components of the
super-curvatures and torsion
and are treated in a completely symmetrical fashion: therefore in our
formulation the ``self-dual'' nature of $D=10$, $N=1$ SUGRA is manifest.

The constraint on the supercurvature, mentioned above, which we introduce
consists in setting to zero the spinorial components of the supercurvature
two-form $R_c{}^d= {1\over 2} E^B E^A R_{ABc}{}^d$,

$$
R_{\alpha\beta ab}=0\au{1}
$$

\noindent
as suggested in \cite{PDA}. Here $A$ indicates both a vectorial index
$a$ and a spinorial index $\alpha$. The constraint \e{1} resembles much
the algebraic structure of the Super-Yang--Mills theory (SYM) in ten
(and also in other) dimensions. We recall in fact, that if we indicate
with $F={1\over 2} E^A E^B F_{BA}$ the Lie algebra valued
Yang--Mills curvature
two-form the constraint $F_{\alpha\beta}=0$ sets the theory on shell
(in $D=10$) in that it implies the equations of motion for gluons and
gluini. As we will see, precisely the same happens also for $N=1$, $D=10$ pure
SUGRA: the constraint \e{1} imposes all the equations of motion for the
supergravity fields and implies, moreover, the existence of a closed
three-superform and  of a closed seven-superform. We would like to remember
that
this analogy between SUGRA and SYM holds only for pure SUGRA in that, if
one constructs non minimal models e.g.
coupling the supergravity to gauge fields, the constraint \e{1} can no
longer be imposed \cite{PDA}.

A remarkable advantage of having a supercurvature two-form satisfying
\e{1}
results from the following considerations regarding anomalies. As is known
$N=1$, $D=10$ pure SUGRA is plagued by an ABBJ Lorentz anomaly $A_L$ due to
the fact that the theory contains chiral fermions and that $D/2+1$
is even. The Lorentz anomaly $A_L$ can be computed via standard techniques
through the so-called extended transgression formula \cite{ETF} starting
from the twelve-form

$$
X_{12} = {62\over945}\ tr\ R^6 - {7\over180}\ tr\ R^4\ tr\ R^2 + {1\over216}
\ (tr\ R^2)^3
\au{2}
$$

\noindent
where with $R_a{}^b$ we
mean here the curvature two-form in {\it ordinary} space. The procedure to
compute $A_L$ relies heavily on the following properties of $X_{12}$:
it is Lorentz-invariant, closed $dX_{12}=0$, and it vanishes being a
twelve-form in ten dimensions. If we indicate with $\Omega_L$ the
BRST operator associated to Lorentz transformations, $A_L$ satisfies the
Wess--Zumino consistency condition:

$$
\Omega_L A_L = 0.\au{3}
$$
It is however clear that $A_L$, being the standard ABBJ-anomaly, is not
supersymmetric. If we indicate with $\Omega_S$ the BRST operator
associated with supersymmetry $\Omega_S A_L\not = 0$, meaning that there
is also a non vanishing SUSY-anomaly $A_S$ in the theory. Therefore one
has to cope with the following coupled cohomology problem \cite{WANOM}:

$$\eqalign{
\Omega_L   A_L & =0\cr
\Omega_S A_L + \Omega_L A_S & =0\cr
\Omega_S A_S  & =0.\cr}\au{4}
$$
A straightforward extension of the transgression method, which allowed
to determine $A_L$ in ordinary space, to superspace is not available
because as a {\it superform} $X_{12}$ does not vanish in $D=10$ superspace.
However, in \cite{PDA} it has been shown that an explicit solution
of the coupled cohomology problem \e{4} can be given provided the
superform $X_{12}$ satisfies ``Weyl triviality'', i.e., there exists
a {\it Lorentz-invariant} eleven-superform $Y$ such that

$$
X_{12} = dY.\au{5}
$$
Note that there exists always an eleven-superform $Y_{CS}$ of
the Chern--Simons type, simply due to the fact that $X_{12}$ is closed;
$dX_{12} = 0\ \Rightarrow\ X_{12} = dY_{CS}$, but $Y_{CS}$
is not Lorentz-invariant. On the other hand, it can be shown
\cite{PDA} that Weyl triviality \e{5} holds provided the constraint
$R_{\alpha\beta a b}=0$ is satisfied.

We conclude that in the present formulation the coupled cohomology
problem \e{4} can be explicitly solved and an explicit expression for the
supersymmetric partner $A_S$ of the Lorentz-anomaly can be determined,
in complete analogy with the SYM theory in ten dimensions.

\vskip0.3truecm
The issue of duality in $N=1$, $D=11$ SUGRA gets settled in a somehow
different manner. The physical content of this theory is given by the
graviton $E_m{}^a$, the gravitino $E_m{}^\alpha$ and by additional bosonic
degrees of
freedom which, at the kinematical level, can be described by a three-form
potential $B_3$ or a six-form potential $B_6$, suggesting
a duality relation between
the field strengths
$H_4=dB_3$ and $\tilde H_7=dB_6$.

Also in this case we reformulate the theory in superspace in a strictly
super-geometrical framework without introducing any closed $H_4$ or
$\tilde H_7$ at the
beginning. This time the theory is put on shell by setting to zero
a certain eleven-dimensional spinor superfield while in eleven dimensions
$R_{\alpha\beta ab}$ remains intrinsically non vanishing in that it
can not be eliminated by any field redefinitions. The Bianchi identities on the
torsion imply then the existence of a 4-superform $H_4$ and of a
7-superform $\tilde H_7$ such that

$$
dH_4=0\au{6}
$$

$$
d \tilde H_7=0.\au{7}
$$
This means that $N=1$, $D=11$ SUGRA is self-dual from a super-kinematical
point of view, but we will see that this self-duality is broken at a
dynamical level, as it is well known in the literature from many years
\cite{TOR}. This fact agrees of course also with the observation that there
exists a super two-brane which lives in an $N=1$, $D=11$ SUGRA background, but
that no dual $p$-brane, living in the same background, is known to exist.

\vskip0.3truecm
The paper is organized as follows. In section two we discuss the general
framework of our formulation of $N=1$, $D=10$ pure SUGRA. In section three
we solve the Bianchi identities. In section four we determine the
equations of motion and evidenciate the self-dual structure of the theory.
Sections five and six are devoted to $N=1$, $D=11$ SUGRA while in section
seven we discuss,
in the present formulation, non minimal theories in ten and eleven
dimensions.
A technical
appendix containing our conventions and
some group-theoretical considerations on $SO(10)$ and
$SO(11)$ concludes the paper.

\autosez{II} The structure of pure supergravity in
ten dimensions\par

The $N=1$, $D=10$ pure supergravity \cite{CHAM} multiplet is given by the
graviton $E_m{}^a$, the chiral gravitino $E_m{}^\alpha$, the dilaton $\phi$,
a chiral fermion which we call gravitello $V_\alpha$ and by additional 28
bosonic degrees of freedom which can be described either by a 2-form
potential $B_{a_1 a_2}$ or by a 6-form potential $B_{a_1-a_6}$
\cite{CHD}.

A superspace in ten dimensions \cite{NI} is spanned by the coordinates
$z^M=(x^m,\theta^\mu)$ where $x^m$ $(m=0,1,\ldots,9)$ are the ordinary
space-time coordinates and $\theta^\mu$ $(\mu=1,\ldots,16)$ are Grassmann
variables. We introduce the supervielbein one-forms $E^A=dz^M E_M{}^A
(z)$ where $A=\{a,\alpha\}$ $(a=0,1\ldots,9;\ \alpha=1,\ldots,16)$ is a flat
index
(letters from the beginning of the alphabet represent flat indices: small
latin letters indicate vectorial indices, small greek letters indicate
spinorial indices and capital letters denote both of them).
The $p$-superforms can be decomposed in the vielbein basis as

$$
\phi_p = {1\over p!} E^{A_1}-E^{A_p} \phi_{A_p-A_1}(z).
$$
We denote the Lorentz-valued super- spin connection one-form by
$\Omega_A{}^B=dz^M \Omega_{M A}{}^B=E^C\Omega_{CA}{}^B$ and the
corresponding covariant differential is written as $D$, while $d$
indicates the ordinary superspace differential. A superfield
$\psi_A{}^B$ is said to be Lorentz-valued if $\psi_{ab}=-\psi_{ba}$ and
$\psi_\alpha{}^\beta={1\over 4} (\Gamma^{ab})_\alpha{}^\beta\psi_{ab}$. Here
we defined

$$
\Gamma^{a_1-a_k} \equiv \Gamma^{[a_1}-\Gamma^{a_k]}
$$
and the matrices $(\Gamma^a)_{\alpha\beta}$ and $(\Gamma_a)^{\alpha\beta}$
are Weyl matrices satisfying the Weyl algebra (see the appendix)

$$
(\Gamma^a)_{\alpha\beta} (\Gamma^b)^{\beta\gamma} +
(\Gamma^b)_{\alpha\beta} (\Gamma^a)^{\beta\gamma} = 2\eta^{ab}
\delta^\gamma_\alpha.
$$
The torsion two-form and the Lorentz-valued curvature two-form are
defined respectively as
$$\eqalign{
T^A & = DE^A = {1\over 2} E^B E^C T_{CB}{}^A\cr
R_A{}^B & = d\Omega_A{}^B + \Omega_A{}^C \Omega_C{}^B = {1\over 2}
E^C E^D R_{DCA}{}^B\cr}
\au{8}
$$
and satisfy the Bianchi identities
$$
DT^A = E^B R_B{}^A\au{9}
$$
$$
DR_A{}^B = 0.\au{10}
$$
Notice that we do not introduce any two- or six-form potential.

The above introduced superfields contain a huge number of unphysical
fields which have to be eliminated by imposing suitable constraints
on the torsion $T_{AB}{}^C$ and on the curvature $R_{ABC}{}^D$. Once
constraints are imposed the Bianchi identities are no longer identities and
they have to be solved consistently.

As it has been shown in \cite{DRA} once the torsion Bianchi identities \e{9}
are consistently solved the Bianchi identities for the curvature \e{10} are
automatically satisfied. This implies that it is sufficient to solve the
torsion Bianchi identities which in components read as:

$$
D_{[A} T_{BC)}{}^D + T_{[AB}{}^G
T_{GC)}{}^D = R_{[ABC)}{}^D\au{10a}
$$

\noindent
where the symbol $[\cdots)$ indicates graded symmetrization. In what
follows $[\cdots]$ will denote antisymmetrization and $(\cdots)$ symmetrization
of
indices.

Our  starting point is the fundamental rigid supersymmetry preserving
constraint

$$
T_{\alpha\beta}{}^a = 2\Gamma^a_{\alpha\beta}.\au{11}
$$
\noindent
Starting from this constraint we can simplify the other components of
the torsion via field redefinitions through the following considerations.
In terms of irreducible representations (irrep) of $SO(10)$ we can
decompose $T_{\alpha\beta}{}^\gamma$ and $T_{\alpha a}{}^b$ as follows:

$$
T_{\alpha\beta}{}^\gamma = (1440\oplus560\oplus144\oplus2\cdot16)\au{12}
$$
$$
T_{\alpha a}{}^b = (720\oplus560\oplus2\cdot144\oplus2\cdot16)\au{13}
$$
\noindent
Through the field redefinitions \ccite{PDV}{ST}

$$
\eqalign{
E^{'\alpha} &= E^\alpha + E^b\,H_b{}^\alpha \cr
\Omega'_{\alpha a}{}^b &= \Omega_{\alpha a}{}^b + X_{\alpha a}{}^b,\cr
}
\au{13a}
$$

\noindent
where $H_b{}^\alpha$ and $X_{\alpha a}{}^b$ are suitable covariant
superfields, we can eliminate from $T_{\alpha a}{}^b$ all the irreps apart
from the 720. Writing now \e{10a} in the lowest sector

$$
(\Gamma^a)_{\delta(\alpha} T_{\beta\gamma)}{}^\delta =
(\Gamma^g)_{(\alpha\beta} T_{\gamma)g}{}^a\au{14}
$$

\noindent
and noting that according to \e{12} $T_{\alpha\beta}{}^\gamma$ does not
contain the irrep 720 also $T_{\alpha a}{}^b$ can not contain it and
must therefore vanish. Noting that the general content of irreps
in \e{14} is $5280\oplus1440\oplus720\oplus560\oplus144\oplus16$ and that the
r.h.s. of \e{14} is zero, we
conclude that $T_{\alpha\beta}{}^\gamma$ can contain only an irrep
16, which corresponds to a spinor $V_\alpha$. A short calculation gives
then

$$
T_{\alpha\beta}{}^\gamma = 2 \delta^\gamma_{(\alpha} V_{\beta)}-
(\Gamma^g)_{\alpha\beta}(\Gamma_g)^{\gamma\varphi} V_\varphi.\au{15}
$$
\noindent
All these considerations were of purely kinematical nature.

We introduce now dynamics by imposing that the purely spinorial
components of the supercurvature vanish:

$$
R_{\alpha\beta ab} =0\,;\au{16}
$$

\noindent
we will in fact see that with this constraint the theory is set on shell.

To summarize, our basic parametrizations for supercurvature and torsion
are

$$\eqalign{
T_{\alpha\beta}{}^a &= 2\Gamma^a_{\alpha\beta}\cr
T_{\alpha\beta}{}^\gamma & = 2\delta_{(\alpha}^\gamma V_{\beta)}-
(\Gamma^g)_{\alpha\beta}(\Gamma_g)^{\gamma\varphi} V_\varphi\cr
T_{\alpha a}{}^b & = 0 = T_{a \alpha}{}^b\cr
R_{\alpha\beta a b} & = 0.\cr}\au{17}
$$
\noindent
In the next section we will see that the closure of the superalgebra
implies a constraint on the superfield $V_\alpha$ which can be identically
solved if one says that $V_\alpha$ is the spinorial derivative of a
scalar superfield, the dilaton $\phi$

$$
V_\alpha = D_\alpha \phi.\au{18}
$$

\noindent
We would like to stress that, without any additional assumption, \e{17}
and \e{18} are sufficient  to determine the theory completely and to
imply in particular all the equations of motion demanding the closure
of the SUSY-algebra, as will be seen in the next section. Under
``closure of the SUSY-algebra" we understand the consistency of the
Bianchi identities with the commutator algebra:

$$
D_A D_B-(-)^{AB} D_B D_A = - T_{AB}{}^C D_C -
R_{AB\#}{}^\#.\au{19}
$$

\autosez{III} Solution of the torsion Bianchi identities\par

The Bianchi identities \e{10a} which have to be solved are written more
explicitly as follows (the one with the lowest dimensions, see \e{14},
has already been solved):

$$
2T_{a(\alpha}{}^\gamma (\Gamma^b)_{\beta)\gamma} + (\Gamma^g)_{\alpha\beta}
T_{ga}{}^b =0\au{20}
$$

$$
D_{(\alpha} T_{\beta\gamma)}{}^\delta - 2 (\Gamma^g)_{(\alpha\beta}
T_{\gamma)g}{}^\delta + T_{(\alpha\beta}{}^\varepsilon
T_{\gamma)\varepsilon}{}^\delta
=0\au{21}
$$

$$
D_\alpha T_{ab}{}^c + 2 (\Gamma^c)_{\alpha\gamma} T_{ab}{}^\gamma =
2R_{\alpha[ab]}{}^c\au{22}
$$

$$
D_a T_{\alpha\beta}{}^\gamma + 2D_{(\alpha} T_{\beta)a}{}^\gamma +
2T_{a(\alpha}{}^\delta T_{\beta)\delta}{}^\gamma
+2 \Gamma_{\alpha
\beta}^gT_{ga}{}^\gamma
-T_{\alpha\beta}{}^\varepsilon T_{a\varepsilon}{}^\gamma=
2R_{a(\alpha\beta)}{}^\gamma
\au{23}
$$

$$
D_{[a}T_{bc]}{}^d - T_{[ab}{}^g T_{c]g}{}^d = R_{[abc]}{}^d\au{24}
$$

$$
D_\alpha T_{ab}{}^\beta + 2D_{[a}T_{b]\alpha}{}^\beta -
2T_{\alpha[a}{}^\delta T_{b]\delta}{}^\beta + T_{ab}{}^g
T_{g\alpha}{}^\beta + T_{ab}{}^\delta T_{\delta\alpha}{}^\beta =
R_{ab\alpha}{}^\beta\au{25}
$$

\noindent
we want now to present the (unique) solution of \e{20}-\e{25} in
compatibility with \e{17} and \e{18}.

The solution of these equations can be most easily achieved using
group theoretical considerations: every tensor appearing in the
equations gets first decomposed in irreps of $SO(10)$, then the
general content of irreps of each equation has to be established and
finally the equations are solved in each sector of $SO(10)$ irreps
separately; this procedure reduces the necessary $\Gamma$-matrix
gymnastic to a minimum.

Eqs. \e{20} and \e{21} are solved as follows: they imply that the vectorial
torsion $T_{ab}{}^c$
is completely antisymmetric in its three indices and corresponds thus
to a 120 irrep of $SO(10)$:

$$
T_{abc}\equiv T_{ab}{}^d \eta_{dc} =
T_{[abc]};\au{26}
$$

\noindent
moreover,

$$
T_{a \alpha}{}^\beta = {1\over 4} (\Gamma^{bc})_\alpha{}^\beta
T_{abc}\au{27}
$$

$$
D_\alpha D_\beta \phi=-\Gamma_{\alpha\beta}^g D_g \phi +
V_\alpha V_\beta + {1\over 12} (\Gamma_{abc})_{\alpha\beta}
T^{abc}\au{28}
$$

\noindent
(remember that $V_\alpha = D_\alpha\phi$).
These relations represent the unique solution to \e{20} and
\e{21}.

Eqs. \e{22} and \e{23} are solved by the following relations:

$$
D_\alpha T_{abc} = -6 T_{[ab}{}^\varepsilon
(\Gamma_{c]})_{\varepsilon\alpha}\au{29}
$$

$$
R_{a\alpha bc}=2(\Gamma_a)_{\alpha\varepsilon} T_{bc}^\varepsilon\au{30}
$$

$$
(\Gamma^b)_{\alpha\varepsilon} T_{ba}{}^\varepsilon = D_a V_\alpha + {1\over 4}
T_{abc} (\Gamma^{bc})_\alpha{}^\beta V_\beta.
\au{31}
$$

\noindent
Eq. \e{30} says that the curvature with one vector- and one spinor-like
index is proportional to the field strength of the gravitino,
$T_{ab}{}^\alpha$, in analogy with the Yang--Mills case \cite{ADR}
where one gets

$$
F_{a\alpha} = (\Gamma_a)_{\alpha\varepsilon}\chi^\varepsilon
$$

\noindent
where $\chi^\varepsilon$ is the gluino superfield. \e{31} is the equation
of motion for the gravitino.

Eq. \e{24} is the purely vectorial Bianchi identity for a curvature with
torsion and has thus not to be ``solved". It implies in particular that
the antisymmetric part of the Ricci tensor $R_{ab}\equiv R_{acb}{}^c$ is
non vanishing,

$$
R_{[ab]} = - {1\over 2} D^c T_{cab}.\au{32}
$$

\noindent
Eq. \e{25} has to be regarded as an equation which determines the spinorial
derivative of $T_{ab}{}^\beta$, i.e. the supersymmetry transformations
law for the gravitino field strength.

In the next section we will enforce the closure of the SUSY-algebra
via \e{19} to derive the equations of motion and to prove the
self-dual character of the theory.

\autosez{IV} The equations of motion: duality as an outcome\par

So far we have only obtained the equation of motion for the gravitino,
\e{31}. The equation of motion for the graviton can be obtained
contracting \e{25} with $(\Gamma^a\Gamma_c)_\beta{}^\alpha$ and using
in the first term on the r.h.s. the gravitino equation \e{31}. One gets:

$$
R_{bc} = 2 D_c D_b \phi - D_a T^a{}_{bc},\au{33}
$$

\noindent
and symmetrizing this one obtains Einstein's equations

$$
R_{(ab)} = 2 D_{(a} D_{b)} \phi.\au{34}
$$
\noindent
To obtain the equation of motion for the gravitello $V_\alpha$ one has
to work a little bit harder. In the conventional formulations, see
for example \cite{NI}\cite{ADR}\cite{PDV1}, this equation is obtained
demanding the existence of a closed three-superform, suitably
constrained. In the present case we can obtain it by imposing the
closure of the SUSY-algebra on \e{28}. We compute

$$
D_{(\varepsilon} D_{\alpha)} D_\beta \phi = - \Gamma^g_{\varepsilon\alpha}
D_g D_\beta\phi - {1\over 2} T_{\varepsilon\alpha}{}^\delta
D_\delta V_\beta\au{35}
$$

\noindent
and equate this expression to the one obtained applying $D_\varepsilon$
to \e{28}. The net result we get is the equation of motion for the
gravitello:

$$
(\Gamma^a)^{\alpha\beta} D_a V_\beta = 2 (\Gamma^a)^{\alpha\beta}
V_\beta D_a\phi - {1\over 12} (\Gamma_{abc})^{\alpha\beta}
T^{abc} V_\beta.\au{36}
$$
\noindent
The equation of motion for the dilaton follows now as usual by applying
$D_\alpha$ to this equation:

$$
D^a D_a \phi = 2 D_a \phi D^a \phi - {1\over 12}
T_{abc} T^{abc}.\au{37}
$$
\vskip1truecm
\noindent{\it The reconstruction of $H_3$}\par
\vskip0.7truecm\noindent
Now we come to the ``reconstruction" of the equations for the
gravi-photon. We want first construct a closed three-form.
To do this we compute

$$
D_{(\beta} D_{\alpha)} T_{abc} = - \Gamma^g_{\alpha\beta}
D_g T_{abc} - {1\over 2} T_{\alpha\beta}{}^\delta D_\delta
T_{abc}\au{38}
$$

\noindent
and equate this expression to the one obtained applying $D_\beta$ to
\e{29} and using on the r.h.s. again \e{25}. One gets, upon projecting
with $(\Gamma_d)^{\alpha\beta}$,

$$
D_d T_{abc} -6 D_{[a} T_{bc]d}+3R_{[abc]d}-9
T^2_{[abc]d} =0\au{39}
$$

\noindent
where we defined $T^2_{abcd}\equiv T_{ab}{}^g T_{gcd}$. Comparing this
with the Bianchi identity \e{24} we obtain precisely what we need:

$$
D_{[a} T_{bcd]} + {3\over 2}\ T^2_{[abcd]} = 0\au{40}
$$
In fact, if we define now a three-superform
$H_3={1\over 3!} E^A E^B E^C H_{CBA}$ through:

$$\eqalign{
H_{\alpha\beta\gamma} &=0= H_{ab\alpha}\cr
H_{a\alpha\beta} &= 2\ (\Gamma_a)_{\alpha\beta}\cr
H_{abc} &= T_{abc}\cr}
\au{40a}
$$

\noindent
the relations \e{29}, \e{40} and the cyclic identity
$(\Gamma^a)_{(\alpha\beta}(\Gamma_a)_{\gamma)\delta}=0$, imply then that

$$
dH_3=0\au{41}
$$
and therefore
$$
H_3 = d B_2\au{42}
$$
for some two-superform $B_2$ (throughout this paper we assume that there
are no topological obstructions and so all closed forms are also exact).
We conclude that we can interpret
$T_{abc}$ as the curl of a two-form potential, and its equation of
motion can be read off from \e{32} and \e{33}:

$$
D_c T^c{}_{ab} = - 4D_{[a} D_{b]} \phi.\au{43}
$$
\vskip1truecm
\noindent{\it The reconstruction of $H_7$}\par
\vskip0.7truecm\noindent
It is a little bit less straightforward to construct a closed
seven-superform starting from \e{43}. We proceed through the
following steps. First we observe that we can rewrite \e{43} as

$$
D_c\Bigl( e^{-2\phi} T^c{}_{ab}\Bigr)=2
e^{-2\phi} T_{ab}{}^\alpha V_\alpha.\au{44}
$$
\noindent
Defining now a 120 irrep as $V_{abc}=(\Gamma_{abc})^{\alpha\beta}
V_\alpha V_\beta$ we can use the gravitino and gravitello equations
of motion to obtain its divergence as

$$
D^c\Bigl(e^{-2\phi}V_{abc}\Bigr) = e^{-2\phi}
\Bigl((\Gamma^{hg}{}_{ab})_\varepsilon{}^\beta T_{hg}{}^\varepsilon V_\beta
+2 T_{ab}{}^\alpha V_\alpha + T^{c_1 c_2}{}_{[a} V_{b]c_1 c_2}\Bigr).
\au{45}
$$
Eliminating now the term $T_{ab}{}^\alpha V_\alpha$ between \e{44} and
\e{45} we get
$$
D^c\Bigl(e^{-2\phi}(T_{abc} - V_{abc})\Bigr)  = -e^{-2\phi}
\Bigl( (\Gamma^{hg}{}_{ab})_\varepsilon{}^\beta T_{hg}{}^\varepsilon V_\beta
+ T^{c_1 c_2}{}_{[a} V_{b]c_1 c_2}\Bigr).
\au{46}
$$
With the definitions

$$
\eqalign{
H_{a_1-a_7} &\equiv {1\over 3!}\ \varepsilon_{a_1-a_7 b_1 b_2 b_3} e^{-2\phi}
\Bigl( T^{b_1 b_2 b_3}-V^{b_1 b_2 b_3}\Bigr)\cr
H_{\alpha a_1-a_6} &\equiv -2 e^{-2\phi}
(\Gamma_{a_1-a_6})_\alpha{}^\beta V_\beta\cr}
\au{46a}
$$

\noindent
\e{46} can be recast into
$$
D_{[a_1}H_{a_2-a_8]} + {7\over 2}\ T_{[a_1 a_2}{}^b H_{a_3-a_8]b}
+ {7\over 2}\ T_{[a_1 a_2}{}^\delta H_{\delta a_3-a_8]} =0.\au{48}
$$
If we now define, moreover,

$$\eqalign{
H_{\alpha\beta a_1-a_5} &\equiv -2 e^{-2\phi}
(\Gamma_{a_1-a_5})_{\alpha\beta}\cr
H_{\alpha_1 \alpha_2 \alpha_3 a_1-a_4} & =\cdots=
H_{\alpha_1 \alpha_2...\alpha_7} =0\cr}
\au{49}
$$

\noindent
it is a simple (but lengthy) exercise to show that the seven-superform

$$
H_7 \equiv {1\over 7!} E^{A_1}-E^{A_7} H_{A_7-A_1}\au{50}
$$

\noindent
satisfies identically

$$
dH_7 = 0.\au{51}
$$
Equation \e{48} is clearly the projection of \e{51} on the  purely
vectorial sector. Thus we proved the existence of a
six-superform $B_6$ such that

$$
H_7 = d B_6.
$$
The theory admits therefore a double interpretation: we can regard
$T_{abc}$ as a closed three-superform whose equation of motion is
given by \e{43}; otherwise we can express $T_{abc}$ through the first
equation of
\e{46a},
in terms of a closed  seven-form $H_{a_1-a_7}$ whose equation of motion
can be read off directly from \e{40}:

$$\eqalign{
D_b H^b &{}_{a_1-a_6} = -2 D_g\phi\ H^g{}_{a_1-a_6}-6
V_{b_1 b_2[a_1} H^{b_1 b_2}{}_{a_2-a_6]} \cr
+ {1\over 3!} & \varepsilon_{a_1-a_6}{}^{b_1-b_4} e^{-2\phi}
\left(D_{b_1} V_{b_2-b_4}+ {3\over 2}\ V^2_{b_1 b_2 b_3 b_4}\right)\cr
- {1\over 180} & e^{2\phi} \varepsilon_{a_1-a_6 b_1-b_4}
H^{b_1 b_2}{}_{f_1-f_5} H^{b_3 b_4 f_1-f_5}\cr}\au{52}
$$
where $V^2_{abcd}=V_{ab}{}^gV_{gcd}$.

It is worthwhile to notice that \e{40} and \e{52} simplify naturally
if one introduces the torsion-free covariant derivative, $\tilde D_a$,
which is defined in terms of the torsion-free connection
$$
\tilde\Omega_{ab}{}^c = \Omega_{ab}{}^c - {1\over 2}
T_{ab}{}^c.\au{52a}
$$

\noindent
\e{40} and \e{52} become then simply:

$$\eqalign{
\tilde D_{[a}  T_{bcd]} &= \tilde D_{[a}
H_{bcd]}=0\cr
\tilde D_g(e^{2\phi} H^g{}_{a_1-a_6}) &= {1\over 3!}
\varepsilon_{a_1-a_6}{}^{b_1-b_4}
\tilde D_{b_1} V_{b_2 b_3 b_4}.\cr}
$$
\noindent
Notice, however, that the shift \e{52a} would introduce a
non vanishing $R_{\alpha\beta ab}$, as it can easily be seen, and therefore
we did not perform this shift.

It is important to notice that the fundamental duality relation
\e{46a} involves only tensors which are invariant under the gauge
transformations

$$\eqalign{
B_6 & \rightarrow B_6 + d\phi_5\cr
B_2 & \rightarrow B_2 + d\phi_1\cr}
$$

\noindent
where $\phi_{1,5}$ are arbitrary superforms (contrary to what happens
in $N=1$, $D=11$ SUGRA, as we will see). This implies definitely that
(self)-duality holds also at the dynamical level, meaning that one
can write a gauge invariant action and gauge invariant equations of
motion in which appears only $H_7$, or a gauge invariant action and
gauge invariant equations of motion in which appears only $H_3$.

The analysis of the self-duality property in non minimal
$D=10$  supergravity theories will be developed in section seven.

\autosez{V} N=1, D=11 Supergravity: a group theoretical analysis of the
constraints\par

We want derive in the following two sections the superspace equations of
motion of $N=1$, $D=11$ SUGRA according to a strategy analogous to the one
used in the preceding sections to evidenciate the self-duality nature of
$N=1$, $D=10$ pure SUGRA. Through an exhaustive group theoretical analysis of
possible constraints we want also to show in which direction one has to move
if one wants to construct non minimal $N=1$, $D=11$ supergravity
theories. Such non minimal theories are interesting in that they can take
quantum corrections to the classical super two-brane $\sigma$-model into
account, supposed that the super two-brane is a
consistent theory also at the quantum level.

We take here the conservative point of view demanding that the zero-dimension
component of the torsion is the rigid one:
$$
T_{\alpha\beta}{}^a=2\Gamma^a_{\alpha\beta}
$$
(for conventions about $\Gamma$-matrices and notations, see the appendix).
Our starting points are again the Bianchi identities \e{9} and \e{10},
Dragon's theorem holds also here and we have thus to find a consistent
solution of \e{10a}.

We remember that the $N=1$, $D=11$ SUGRA multiplet is made out of the
graviton, $E_m{}^a$, the gravitino $E_m{}^\alpha$ and additional bosonic
degrees of freedom which ``numerically'' can be described in terms of a
three-form $B_3$ or a six-form $B_6$ potential. Also here we do not introduce
any closed four- or seven-superform a priori, but try to reconstruct them in
{\it superspace} by solving solely \e{10a}.

To begin with, we apply the same kinematics as in section two. The
decompositions in terms of $SO(11)$ irreps, analogous to \e{12} and \e{13},
are now
$$
T_{\alpha\beta}{}^\gamma=5280\oplus4224\oplus3520\oplus2\cdot1408
\oplus3\cdot320\oplus3\cdot32
\au{53}
$$
$$
T_{\alpha a}{}^b=1760\oplus1408\oplus2\cdot320\oplus2\cdot32.\au{54}
$$
Precisely as in section two through the field redefinitions \e{13a} we can now
eliminate from $T_{\alpha a}{}^b$ all irreps apart from the 1760. The lowest
order Bianchi identity is formally identical to \e{14}:
$$
(\Gamma^a)_{\delta(\alpha}T_{\beta\gamma)}{}^\delta=
(\Gamma^g)_{(\alpha\beta}T_{\gamma)g}{}^a.
\au{55}
$$
Again the 1760 irrep is not contained in $T_{\alpha\beta}{}^\gamma$ and so
\e{55} implies that $T_{\alpha a}{}^b$ vanishes. Moreover, the general content
of \e{55} is given by
$$
36960\oplus10240\oplus5280\oplus4224\oplus3520\oplus1760
\oplus2\cdot1408\oplus3\cdot320\oplus2\cdot32;
$$
taking a look at \e{53} and noticing that the r.h.s. of \e{55} vanishes, we
conclude that all irreps of $T_{\alpha\beta}{}^\gamma$ have to vanish, apart
from {\it one} 32 (the spinorial representation). This is due to the fact that
$T_{\alpha\beta}{}^\gamma$ contains three 32 irreps and that \e{55} establishes
two linear relations among them; therefore only one of them is independent.
A short computation gives then
$$
T_{\alpha\beta}{}^\gamma=16\delta_{(\alpha}^\gamma
V_{\beta)}-6(\Gamma^g)_{\alpha\beta}(\Gamma_g)^{\gamma\delta}V_\delta+
(\Gamma^{ab})_{\alpha\beta}(\Gamma_{ab})^{\gamma\delta}V_\delta
\au{57}
$$
where we identify $V_\alpha$ as the independent 32 irrep. This relation
substitutes eq. \e{15} in ten dimensions. A more fundamental difference
between $D=10$ and $D=11$ SUGRA comes in at this point: the theory is set on
shell and reduces to {\it pure} $N=1$, $D=11$ SUGRA if we set (as a dynamical
constraint):
$$
T_{\alpha\beta}{}^\gamma=0\ \Leftrightarrow\ V_\alpha=0.
\au{58}
$$
As we will see $R_{\alpha\beta ab}$ is in this case {\it intrinsically\/}
different from zero, it can not be set to zero by field redefinitions.
So with respect to the ten-dimensional case the situation is completely
reversed: there we could set $R_{\alpha\beta ab}=0$ and then
$T_{\alpha\beta}{}^\gamma$ survived, here it is precisely the opposite!

Once we have established that $R_{\alpha\beta ab}$ is different from zero we
can shift the vectorial connection $\Omega_{ab}{}^c$ to set $T_{ab}{}^c$
to zero (notice that in $D=10$ the analogous shift in the connection was not
performed in that it would have turned on $R_{\alpha\beta ab}$).

To conclude, $D=11$, $N=1$ pure supergravity can be derived through the
following constraints
$$
\eqalign{
T_{\alpha\beta}{}^a&=2\Gamma^a_{\alpha\beta}\cr
T_{\alpha a}{}^b&=T_{\alpha\beta}{}^\gamma=T_{ab}{}^c=0.\cr}
\au{59}
$$

What we learned mainly from all this is that, within the framework in which
$T_{\alpha\beta}{}^a=2\Gamma^a_{\alpha\beta}$, the {\it unique} way to
construct a non minimal $N=1$, $D=11$ SUGRA (modulo field redefinitions) is to
introduce a non vanishing spinorial 32 irrep in $T_{\alpha\beta}{}^\gamma$. We
will comment on the possible significances of this relaxed constraint and on
the importance the resulting theories would have in section seven. Here we
proceed by rederiving minimal $N=1$, $D=11$ SUGRA relying on group theoretical
reasonings and paying particular attention to the duality structure of the
theory.

The Bianchi identities are very similar to \e{20}-\e{25}. Eqs. \e{20} and
\e{21} are substituted by:
$$
\eqalign{
4T_{a(\alpha}{}^\gamma\Gamma^b_{\beta)\gamma}&=R_{\alpha\beta a}{}^b\cr
2\Gamma^g_{(\alpha\beta}T_{g\gamma)}{}^\delta&=R_{(\alpha\beta\gamma)}
{}^\delta\cr}
\au{60}
$$
(remember that $R_{\alpha\beta\gamma}{}^\delta={1\over4}R_{\alpha\beta
ab}\Gamma^{ab}{}_\gamma{}^\delta$). The remaining Bianchi identities are
obtained from \e{22}-\e{25} by simply enforcing \e{59} and we will not write
them down explicitly.

The group theoretical reasoning which allows one to solve \e{60} is reported
in the appendix. Here we state simply the result.

It turns out that $R_{\alpha\beta ab}$ and $T_{a\alpha}{}^\beta$ are expressed
in terms of {\it one} 330 irrep (which corresponds to a completely
antisymmetric rank four tensor) $W_{abcd}$. One gets
$$
\eqalign{
T_{a\alpha}{}^\beta&=8\ (\Gamma^{b_1b_2b_3})_{\alpha\beta}W_{ab_1b_2b_3}+
(\Gamma_{ab_1-b_4})_{\alpha\beta}W^{b_1-b_4}\cr
R_{\alpha\beta ab}&=96\ (\Gamma^{c_1c_2})_{\alpha\beta}W_{c_1c_2ab}+
4\ (\Gamma_{abc_1-c_4})_{\alpha\beta}W^{c_1-c_4}.\cr}
\au{62}
$$
Equations \e{22} and \e{23} are solved by the relations
$$
R_{\alpha abc}=2T_{a[b}{}^\delta\Gamma_{c]\delta\alpha}
-T_{bc}{}^\delta(\Gamma_a)_{\delta\alpha}
\au{63}
$$
$$
D_\alpha W_{a_1-a_4}={1\over24}(\Gamma_{[a_1a_2})_{\alpha\gamma}
T_{a_3a_4]}{}^\gamma.
\au{63a}
$$
Consistency implies also that among the three irreps contained in
$T_{ab}{}^\alpha$, i.e. $1408\oplus320\oplus32$, only the highest one, i.e.
the 1408, is
non vanishing. This implies immediately the gravitino equation of motion
$$
(\Gamma^{abc})_{\alpha\beta}T_{bc}{}^\beta=0
\au{64}
$$
and that the ``trace'' of $T_{ab}{}^\alpha$ vanishes:
$$
T_{ab}{}^\alpha(\Gamma^b)_{\alpha\beta}=0.
\au{64a}
$$
\autosez{VI} Duality in D=11\par
The identity \e{24}, remembering that now $T_{ab}{}^c=0$, implies that
$R_{c[ab]}{}^c=0$ meaning that the Ricci tensor is symmetric,
$R_{ab}=R_{(ab)}$.
\e{25} instead can be written as
$$
D_\alpha T_{ab}{}^\beta=-2D_{[a}T_{b]\alpha}{}^\beta
-2T_{[a\alpha}{}^\gamma T_{b]\gamma}{}^\beta
+{1\over4}R_{abcd}(\Gamma^{cd})_\alpha{}^\beta.
\au{65}
$$
Using then the tracelessness of $T_{ab}{}^\beta$ and contracting \e{65} with
$(\Gamma^b\Gamma_c)_\beta{}^\alpha$ we obtain Einstein's equations
$$
R_{ab}-{1\over2}\eta_{ab}R=-288\cdot4!\ \left(W^2_{ab}-{1\over8}
\ \eta_{ab}W^2\right)
\au{66}
$$
where we defined $R=R^a{}_a$, $W^2_{ab}\equiv W_{ac_1c_2c_3}W_b{}^{c_1c_2c_3}$,
$W^2\equiv W_{a_1-a_4}W^{a_1-a_4}$.

We compute now
$$
(\Gamma_{a_5})^{\beta\alpha} D_\beta D_\alpha
W_{a_1-a_4}=-32D_{a_5}W_{a_1-a_4}.
\au{67}
$$
The left hand side of this equation can also be evaluated by applying $D_\beta$
to \e{63a} and using eq. \e{65} for $D_\beta T_{a_3a_4}{}^\gamma$.

The net result of this computation is the important relation
$$
D_{[a_1}W_{a_2a_3a_4a_5]}=0.
\au{68}
$$
If we define now
\medskip\noindent
$$
\eqalign{
H_{a_1-a_4}&=W_{a_1-a_4}\cr
H_{ab\alpha\beta}&=-{1\over144}(\Gamma_{ab})_{\alpha\beta}\cr
H_{\alpha\beta\gamma\delta}&=H_{\alpha\beta\gamma a}=H_{\alpha abc}=0\cr}
\au{69}
$$
\medskip\noindent
and then as usual $H_4={1\over4!}E^{A_1}-E^{A_4}H_{A_4-A_1}$, eq. \e{68},
together
with other relations of the present and the preceding section imply that
$H_4$ is a
closed superform
$$
dH_4=0
\au{70}
$$
and therefore we can define a $3$-form superpotential $B_3$ such that
$$
H_4=dB_3.
\au{71}
$$
\bigskip\noindent
Using now again the tracelessness of the gravitino field strength in the
form
$$
D_\alpha T_{ag}{}^\beta(\Gamma^g\Gamma_{bc})_\beta{}^\alpha=0
$$
and substituting \e{65} we get
$$D_dW^d{}_{abc}=-{1\over4}\varepsilon_{abcf_1-f_4g_1-g_4}W^{f_1-f_4}W^{g_1-g_4}
\au{72}
$$
which can be read as the equation of motion for $H_4$.
However, if we define a seven-superform
$H_7={1\over7!}E^{A_1}-E^{A_7}H_{A_7-A_1}$ through
$$
\eqalign{
H_{a_1-a_7}&={1\over4!}\varepsilon_{a_1-a_7b_1-b_4}W^{b_1-b_4}\cr
H_{\alpha\beta a_1-a_5}&={1\over144}\ (\Gamma_{a_1-a_5})_{\alpha\beta}\cr}
\au{73}
$$
and all other components of $H_{A_1-A_7}$ vanishing, then \e{72}, together with
other relations of the last two sections, implies the {\it superspace} relation
$$
dH_7={1\over144}H_4\wedge H_4.
\au{74}
$$
Substituting \e{71} we can write this as
$$
d\left(H_7-{1\over144}B_3\wedge H_4\right)=0
$$
meaning that the seven-superform
$\tilde{H}_7\equiv H_7-{1\over144}B_3\wedge H_4$ is closed,
$$
d\tilde{H}_7=0\qquad\Rightarrow\qquad\tilde{H}_7=dB_6,
\au{75}
$$
for some six-superform $B_6$.
\e{71} and \e{75} imply that in $N=1$, $D=11$ SUGRA duality holds at the {\it
kinematical\/} level in superspace, meaning that one can always construct a
closed four-superform and a closed seven-superform. We observe also that we
can write
$$
H_7=dB_6+{1\over144}B_3\wedge H_4
\au{76}
$$
which resembles much the relation which couples in $N=1$, $D=10$ the
super-Maxwell multiplet to $N=1$, $D=10$ SUGRA \cite{SUSM}:
$$
H_3=dB_2+kA\wedge F
\au{77}
$$
where $A$ and $F$ are the connection $1$-form and curvature $2$-form
respectively. Gauge invariance in \e{77}, $A\rightarrow A+d\phi$, is saved by
demanding that
$$
B_2\rightarrow B_2-k\phi\wedge F.
$$
Similarly we can save gauge invariance in \e{76} by demanding that
$B_3\rightarrow B_3+d\phi_2$ be accompanied by
$$
B_6\rightarrow B_6-{1\over144}\phi_2\wedge H_4.
$$
Thus gauge invariance and duality hold, at a kinematical level, in
superspace.

{}From a dynamical point of view, however, one has to observe that if one
reads the eqs. of motion \e{66} and \e{72} in terms of $H_4$, then the
potential $B_3$ appears obviously in a local way simply because $H_4=dB_3$;
on the other hand those equations can also be interpreted as equations of
motion which involve $H_7$ through local (polynomial) expressions, see
\e{73}, and the equation of motion for $H_7$ would then simply be
(see \e{68})
$$
D^bH_{ba_1-a_6}=0.
\au{78}
$$
However, the relation between $B_6$ and $H_7$ becomes now {\it non local}.
In fact, if one ``inverts'' the relation $H_4=dB_3$ to get a non local
expression for $B_3$ in terms of $H_4$, or equivalently in terms of $H_7$,
\e{76} produces an implicit and non local relation between $H_7$ and $B_6$.
We conclude, therefore, that in the dual interpretation, i.e. in terms of a
closed seven-form, $N=1$, $D=11$ SUGRA becomes non local, as it is already
known in the literature of course, and we are forced to formulate the
theory in terms of a closed four-form.

\autosez{VII} Conclusions and further developments\par
Let us first make some remarks on supergravity theories in ten dimensions.

As we saw, in our approach in the pure supergravity theory a closed
three-superform and a closed seven-superform arise naturally, the unique
dynamical constraint being $R_{\alpha\beta ab}=0$. For, to couple the theory
to e.g. Yang--Mills fields or to construct non minimal couplings in pure
supergravity theories (or both) one has to release this constraint introducing
a non vanishing $R_{\alpha\beta ab}$. On completely general grounds relying
only on the constraint $T_{\alpha\beta}{}^a=2\,\Gamma^a_{\alpha\beta}$, one
can find that the most general parametrization of
$R_{\alpha\beta ab}$, modulo field redefinitions, is in terms of a single
$120$ irrep superfield \cite{PDV}
$$
R_{\alpha\beta ab}=(\Gamma_{abc_1c_2c_3})_{\alpha\beta}J^{c_1c_2c_3}
\au{79}
$$
where the $120$ irrep $J^{abc}$ plays the role of a current. Accordingly,
$T_{a\alpha}{}^\beta$ gets corrected to
$$
T_{a\alpha}{}^\beta={1\over4}(\Gamma^{bc})_\alpha{}^\beta
T_{abc}-{1\over4}(\Gamma_{abcd})_\alpha{}^\beta J^{bcd}.
\au{80}
$$
The solution of the torsion-Bianchi identities with the (most general)
parametrization \e{79} leads to a modification of all equations of motion and
to one constraint on the highest irrep contained in the spinorial derivative
of $J_{abc}$
$$
\left[D_\alpha(e^{2\phi}J_{abc})\right]^{1200}=0.
\au{81}
$$
Once this constraint is satisfied it can be shown that, starting from
$J_{abc}$, one can construct a closed four-superform $K$, such that
$$
dH_3=K
\au{82}
$$
where $H_3$ is again defined as in \e{40a}
and
$$\eqalign{
K_{\alpha\beta\gamma\delta}&=K_{\alpha\beta\gamma a}=0\cr
K_{\alpha\beta ab}&=
2(\Gamma_{abc_1c_2c_3})_{\alpha\beta}J^{c_1c_2c_3}\cr
dK&=0\cr}
\au{82a}
$$
with some more complicated expressions for $K_{\alpha abc}$ and $K_{abcd}$.
Therefore a three-superform $\Omega$ exists such that $K=d\Omega$ and
hence $d(H_3-\Omega)=0$, or
$$
H_3=dB_2+\Omega
\au{83}
$$
for some two-form potential $B_2$.

Similarly one can show that it is also
possible to construct a closed seven-superform $H_7$
$$
dH_7=0
\au{83b}
$$
through
$$
\eqalign{
H_{a_1-a_7}&={1\over3!}\ e^{-2\phi}\varepsilon_{a_1-a_7}{}^{b_1-b_3}
(T_{b_1-b_3}-V_{b_1-b_3}-6J_{b_1-b_3})\cr
H_{\alpha a_1-a_6}&=-2e^{-2\phi}(\Gamma_{a_1-a_6})_\alpha{}^\varepsilon
V_\varepsilon\cr
H_{\alpha\beta a_1-a_5}&=-2e^{-2\phi}(\Gamma_{a_1-a_5})_{\alpha\beta},\cr
}
\au{83h}
$$
while all other components of $H_7$ are vanishing. Remember that, according to
\e{40a}, which holds also in the extended case under investigation in the
present section, $T_{abc}=H_{abc}$.

We conclude that also in the case of $D=10$ extended SUGRA models the theory
can be read in two ways: either \e{83b} is interpreted as the Bianchi identity
for the $B_6$ potential, and then \e{82} is its equation of motion, or
\e{82} is interpreted as the Bianchi identity for $B_2$ through \e{83}, and
then \e{83b} is its equation of motion.

Clearly all this fits precisely in what is known in the literature. In fact,
in order to couple to the SYM fields, one can search for a
decomposition of the type
$$
d\omega_{YM}=TrF^2=dX_{YM}+K_{YM} \au{83a}
$$
where $F={1\over 2}E^AE^BF_{BA}$
is the Lie algebra valued Yang--Mills supercurvature two-form
and $\omega_{YM}$ is the associated Chern--Simons three-superform.
Choosing for $F$ standard constraints, i.e. $F_{\alpha\beta}=0$,
it can easily be shown that \e{83a} holds, with a four-form $K_{YM}$
satisfying \e{82a}, if one chooses for
$X_{YM}$ the gauge-invariant 3-superform
$$
X_{YM}=-{1\over48}E^cE^bE^a(\Gamma_{abc})_{\alpha\beta}
Tr(\chi^\alpha\chi^\beta).
$$
Here $\chi^\alpha$ is the gluino superfield (a Lie algebra valued spinor).

The coupling to the Lorentz Chern--Simons form
is formally analogous; however, now it is not trivial
to show that \cite{PDV1}\cite{BPT1}\cite{TOR3}
$$
d\omega_L=trR^2\equiv R_a{}^bR_b{}^a=dX_L+K_L,
$$
where $X_L$ is a Lorentz invariant 3-form, whose explicit expression is rather
lengthy and can be found e.g. in \ccite{PDV1}{TOR3}, and $K_L$ satisfies again
\e{82a}. In both cases, as it is well known, $B_2$
has to transform anomalously under gauge transformations because $\omega_L$
and $\omega_{YM}$ are not invariant. The invariant superforms $X_{YM}$ and
$X_L$ redefine simply $H_3$, in that $\Omega$ is given by
$\Omega=(\omega_{YM}-\omega_L)-(X_{YM}-X_L)$ (see below).

There is a third case of relevance in the
literature \ccite{LPT}{LP} in which $K$ is the differential of an {\it
invariant} three-superform, called $Z$ in \ccite{LPT}{LP}, which gives rise to
superstring corrections of the minimal pure supergravity which are not
dictated by anomaly cancellation like, for example,
to a term in the action which is proportional to the fourth power of the
Riemann curvature, $(R_{abcd})^4$.

Our main conclusion with respect to $N=1$, $D=10$ supergravity theories is
that, on completely general grounds, relying on the unique (kinematical)
hypothesis that the zero-dimension component of the torsion be flat,
$T_{\alpha\beta}{}^a=2\,\Gamma^a_{\alpha\beta}$, there exist always closed
three and seven-superforms, and that the theory is therefore intrinsically
self-dual; to repeat, this is true for {\it every} non minimal extension of
the theory based on $T_{\alpha\beta}{}^a=2\,\Gamma^a_{\alpha\beta}$.

Finally, we would like to recall that, as we saw, for non minimal
theories $R_{\alpha\beta ab}$ is no longer zero (see \e{79}) and therefore
it is far from obvious that
the twelve superform which triggers the ABBJ Lorentz, gauge and
mixed anomalies \cite{B2AN} satisfies the Weyl triviality property \e{5}
(we remember that this property allows one to compute the supersymmetric
partner of the ABBJ anomaly). We hope to be able to prove Weyl triviality
for this twelve superform in our formulation in that, in contrast to
previous formulations, if one switches off the ``external current''
$J_{abc}$ the extended models reduce to a Weyl trivial model
(pure SUGRA) thanks to $R_{\alpha\beta ab}=0$; this constraint is a
sufficient condition for Weyl triviality to hold, but it should not be
necessary.

The three above mentioned non-minimal extensions of the theory are
relevant in that the resulting extended SUGRA theories describe the low-energy
dynamics of the heterotic superstring. Particularly interesting is the theory
based on the $3$-form field strength
$$
\eqalign{
H_3&=dB_2+(\omega_{YM}-X_{YM})-(\omega_L-X_L)\cr
dH_3&=Tr F^2-tr R^2-d(X_{YM}-X_L)\cr}
\au{84}
$$
which is related to the Green--Schwarz anomaly cancellation mechanism
\cite{B2AN}. Recently it has been argued that the heterotic five-brane whose
background theory is an $N=1$, $D=10$ SUGRA, is dual to the heterotic string
\cite{STRO} and, as a test of this conjecture, it has been shown \cite{DLU}
that the Lorentz and gauge anomalies cancel in the heterotic five-brane via a
mechanism which can be regarded as dual to the Green--Schwarz one. It is based
on a seven-form $H_7$ satisfying (in ordinary ``bosonic'' space)
$$
\eqalign{
dH_7&={1\over24}TrF^4-{1\over7200}(TrF^2)^2-{1\over240}TrF^2trR^2
+{1\over8}trR^4+{1\over32}(trR^2)^2\cr
&\equiv X_8=d\omega_7\cr}
\au{85}
$$
where $\omega_7$ is a generalized Chern--Simons form. As originally the
Green--Schwarz mechanism, also the ``dual'' mechanism \cite{B6AN} based on
\e{85} breaks
supersymmetry. To restore supersymmetry one had to find a consistent solution
of the Bianchi identity \e{85} written in superspace. We hope that our
formulation of $N=1$, $D=10$ SUGRA, which gave us a new general insight into
the self-dual nature of the theory, permits us to answer definitely the
question of the compatibility of \e{85} with supersymmetry. This issue is of
some importance because, if a consistent heterotic five-brane exists,
then a consistent
$N=1$, $D=10$ supergravity, based on \e{85}, must also exist and in this case
one would have (formally) a new theory, the five-brane, describing the same
physics as the
heterotic string. Interesting applications of this equivalence could result
for example from the observation that duality interchanges classical
with quantum corrections and therefore, instead of making a quantum
computation in string theory, one could perform a classical computation in the
five-brane.

\noindent We will discuss the consistency of
\e{85} with supersymmetry elsewhere \cite{CL1}.

Regarding $N=1$, $D=11$ SUGRA we would like to comment briefly on the possible
extensions of the minimal theory based on \e{57} with a non vanishing spinor
$V_\alpha$. It is clear that this spinor has not to be a new field, but must
be a (covariant) function of the fields already present in the theory, and
clearly $V_\alpha$ would have to
satisfy a certain number of constraints coming from the Bianchi identities. If
the extended theory has to be consistent then the torsion Bianchi identities
have to imply the existence of a closed $4$-superform (otherwise the gauge
invariance, needed to eliminate the unphysical degrees of freedom of the
gravi-photon, is missing). The issue of existence of such extended $N=1$,
$D=11$ supergravity theories is of some relevance because the {\it classical}
supermembrane ($p=2$) lives in an $N=1$, $D=11$ minimal supergravity
background and
the fundamental $k$-invariance of the supermembrane $\sigma$-model holds
true classically if the background fields satisfy the equations of motion
of minimal SUGRA. If the $\sigma$-model is consistent also at the quantum
level \cite{PPT} then one can compute the
$k$-anomalies and the requirement of their cancellation could then give
rise to local non minimal supergravity theories. Along these lines proposals
for non minimal SUGRA theories have been made in \cite{PPT} via a
cohomological analysis of
$k$-anomalies in the supermembrane $\sigma$-model. It would be interesting to
find out
if our general framework for non minimal $N=1$, $D=11$ SUGRA, based only on
the rigid SUSY preserving constraint $T_{\alpha\beta}{}^a=
2\ \Gamma^a_{\alpha\beta}$, fits with the extensions proposed in \cite{PPT};
this check, based on a detailed analysis of non minimal models, will be the
subject of a future publication \cite{CL2}.
\bigskip
\noindent{\it Acknowledgements}\par
\vskip0.3truecm\noindent
The authors would like to thank Mario Tonin for numerous invaluable
discussions.
\vfill\eject

\semiautosez{\rm A} Appendix\par
\noindent{\it 1. Ten dimensional gamma-matrix algebra}\medskip\noindent
\def\epsb{{\underline\varepsilon}}
\noindent We use a Majorana--Weyl representation for the Dirac matrices
$\gamma^a$\footnote{(*)}{The $\epsb$ index runs over the full
$32$ components of Dirac spinors, while the other indices are in the chiral
$16$ (lower indices) or $\overline{16}$ (upper indices) irrep of $SO(10)$.}
$$
\eqalign{
(\Gamma^a)_{\alpha\beta}&=(\gamma^a)_\alpha{}^\epsb C_{\epsb\beta}\cr
(\Gamma^a)^{\alpha\beta}&=C^{\alpha\epsb}(\gamma^a)_\epsb{}^\alpha,\cr}
\au{a1}
$$
where $C$ is the (antisymmetric and idempotent) charge conjugation matrix,
characterized by the Weyl algebra
$$
(\Gamma^a)_{\alpha\beta}(\Gamma^b)^{\beta\gamma}+
(\Gamma^b)_{\alpha\beta}(\Gamma^a)^{\beta\gamma}=
2\eta^{ab}\delta_\alpha{}^\gamma
\au{a3}
$$
(here $\eta^{ab}$ is the ``mostly minus'' metric). We define
$$
\Gamma^{a_1\cdots a_k}=\Gamma^{[a_1}\cdots\Gamma^{a_k]}
\au{a4}
$$
that are subjected to the duality property
$$
\Gamma^{a_1\cdots a_k}=\pm(-1)^{{k\over2}(k+1)}{1\over(D-k)!}
\varepsilon^{a_1\cdots a_D}\Gamma_{a_{k+1}\cdots a_D}
\au{a5}
$$
with the minus sign when the first matrix has low spinor indices, and the plus
sign in the other case. Characteristic of ten dimensions is the cyclic
identity
$$
(\Gamma^g)_{(\alpha\beta}(\Gamma_g)_{\gamma)\delta}=0
\au{a6}
$$
which implies its ``dual''
$$
(\Gamma^g)_{(\alpha\beta}(\Gamma_g{}^{a_1\cdots a_4})_{\gamma\delta)}=0.
\au{a7}
$$
\vfill\eject
\vskip0.5truecm
\noindent{\it 2. Eleven dimensional gamma matrix algebra}\medskip\noindent
In eleven dimensions we switch to the ``mostly plus'' metric to avoid the
appearance of explicit ``$i$'' factors in the
formalism. Through the charge conjugation matrix $C$ we define the matrices
$$
\eqalign{
(\Gamma^a)_\alpha{}^\beta&=(\gamma^a)_\alpha{}^\beta\cr
(\Gamma^a)_{\alpha\beta}&=(\gamma^a)_\alpha{}^\varepsilon
C_{\varepsilon\beta}\cr
(\Gamma^a)^{\alpha\beta}&=C^{\alpha\varepsilon}(\gamma^a)_\varepsilon{}^\beta\cr
(\Gamma^a)^\alpha{}_\beta&=C^{\alpha\varepsilon}(\gamma^a)_\varepsilon{}^\lambda
C_{\lambda\beta}.\cr}
\au{mah}
$$
The symmetric matrices are
$$
\Gamma^a,\Gamma^{a_1a_2},\Gamma^{a_1\cdots a_5},
\Gamma^{a_1\cdots a_6},\Gamma^{a_1\cdots a_9},\Gamma^{a_1\cdots a_{10}}
$$
while the antisymmetric ones are
$$
C,\Gamma^{a_1\cdots a_3},\Gamma^{a_1\cdots a_4},
\Gamma^{a_1\cdots a_7},\Gamma^{a_1\cdots a_8},\Gamma^{a_1\cdots a_{11}}.
$$
The duality property becomes
$$
\Gamma^{a_1\cdots a_k}=-(-1)^{{k\over2}(k-1)}{1\over(D-k)!}
\varepsilon^{a_1\cdots a_D}\Gamma_{a_{k+1}\cdots a_D};
\au{a8}
$$
the cyclic identity reads
$$
(\Gamma^{ga})_{(\alpha\beta}(\Gamma_g)_{\gamma\delta)}=0
\au{a9}
$$
and its ``dual'' is now
$$
(\Gamma^g)_{(\alpha\beta}(\Gamma_g{}^{a_1\cdots a_4})_{\gamma\delta)}=
3(\Gamma^{[a_1a_2})_{(\alpha\beta}(\Gamma^{a_3a_4]})_{\gamma\delta)},
\au{a10}
$$
which signals the non vanishing of the curl of $H_7$.
\vskip0.5truecm
\noindent{\it 3. Solution of the dimension one Bianchi identities in D=11}
\medskip\noindent
Here the dimension one Bianchi identities are:
$$
4T_{a(\alpha}{}^\varepsilon(\Gamma_b)_{\beta)\varepsilon}=R_{\alpha\beta ab}
\au{I1}
$$
$$
2(\Gamma^g)_{(\alpha\beta}T_{g\gamma)}{}^\delta=R_{(\alpha\beta\gamma)}
{}^\delta.
\au{I2}
$$
The irrep content of
$T_{a\alpha\beta}=T_{a\alpha}{}^\gamma C_{\gamma\beta}$ is:
$$
\eqalign{
T_{a\alpha\beta}&=T_{a(\alpha\beta)}+T_{a[\alpha\beta]}\cr
&=(11\oplus55\oplus462)\otimes11\oplus(1\oplus165\oplus330)\otimes11\cr
&=4290\oplus3003\oplus1430\oplus2\cdot462\oplus429\cr
&\qquad\oplus2\cdot330\oplus2\cdot165\oplus65
\oplus2\cdot55\oplus2\cdot11\oplus1.\cr
}\au{I4}
$$
Symmetrizing \e{I1} in $(ab)$ and taking into account that $R_{\alpha\beta
ab}$ is antisymmetric in $a,b$ we get
$$
T^{(a}{}_{(\alpha}{}^\varepsilon\Gamma^{b)}_{\beta)\varepsilon}=0.
\au{I3}
$$
The general irrep content of \e{I3} is given by
$$
\eqalign{
(ab)(\alpha\beta)&=(1\oplus65)\otimes(11\oplus55\oplus462)\cr
&=22275\oplus4290\oplus3003\oplus2025\oplus1430\oplus2\cdot462\cr
&\qquad\oplus429\oplus275\oplus65\oplus2\cdot55\oplus2\cdot11\cr}
\au{I5}
$$
so that
$$
T_{a\alpha}{}^\beta=2\cdot330\oplus2\cdot165\oplus1.
\au{I6}
$$
Due to \e{I1} $R_{\alpha\beta a}{}^b$ contains at most the irreps contained
in \e{I6}. Now we can combine eqs. \e{I1} and \e{I2}.
Eq. \e{I2} contains (among a large set of irreps which we are not
interested in) the irreps $3\cdot330\oplus3\cdot165\oplus1$.
By direct inspection one finds that the three linear equations in \e{I2}
involving the two 165 irreps are linearly independent and therefore the two
165 have to vanish, the equation on the singlet implies its vanishing while
the three equations on the two 330 are found to be linearly dependent from
only one of them, meaning simply that the two 330 have to be proportional
to each other.
We conclude that $T_{a\alpha}{}^\beta$ and  $R_{\alpha\beta a}{}^b$ are
made out of a single $330$ irrep (a fourth rank antisymmetric tensor
$W_{abcd}$) in two different forms as shown in \e{62}.
\vfill\eject

\semiautosez{A} References\par
\biblitem{CJS} E.~Cremmer, B.~Julia and J.~Scherk, {\it Phys. Lett.}
{\bf 76B}(1978) 409\par
\biblitem{BH} L.~Brink and P.~Howe, {\it Phys. Lett.} {\bf 91B} (1980) 384;
 E.~Cremmer and S.~Ferrara, {\it Phys. Lett.} {\bf 91B} (1980) 61\par
\biblitem{BST} E.~Bergshoeff, E.~Sezgin and P.~K.~Townsend, {\it Phys. Lett.}
{\bf 189B} (1987) 75\par
\biblitem{CHAM} A.~Chamseddine {\it Nucl. Phys.} {\bf B185} (1981) 403\par
\biblitem{NI} B.~E.~W.~Nillson, {\it Nucl. Phys.} {\bf B188} (1981) 176\par
\biblitem{CHD} A.~Chamseddine {\it Phys. Rev.} {\bf D24} (1981) 3065\par
\biblitem{ACET} A.~Achucarro, J.~M.~Evans, P.~K.~Townsend and D.~L.~Wiltshire,
{\it Phys. Lett.} {\bf B198} (1987) 441\par
\biblitem{STRO} A.~Strominger, {\it Nucl. Phys.} {\bf B343} (1990) 167\par
\biblitem{DLU} M.~J.~Duff and J.~X.~Lu, {\it Phys. Rev. Lett.} {\bf 66} (1991)
1402; {\it Class. Quantum Grav.} {\bf 9} (1991) 1\par
\biblitem{DULU2} M.~J.~Duff and J.~X.~Lu, {\it Nucl. Phys.} {\bf B390} (1993)
276\par
\biblitem{DRA} N.~Dragon, {\it Z. Phys.} {\bf C2} (1979) 29\par
\biblitem{DUFF} M.~J.~Duff, {\it Class. Quantum Grav.} {\bf 5} (1988) 189\par
\biblitem{SUSM} E.~Bergshoeff, M.~De~Roo, B.~De~Wit and P.~Van~Nieuwenhuizen,
{\it Nucl. Phys.} {\bf B195} (1982) 97\par
\biblitem{SUSY} G.~F.~Chapline and N.~S.~Manton, {\it Phys. Lett.} {\bf 120B}
(1982) 105\par
\biblitem{DDP} J.~A.~Dixon, M.~J.~Duff and J.~C.~Plefka, {\it Phys. Rev. Lett.}
{\bf 69} (1992) 3009\par
\biblitem{ADR} J.~J.~Atick, A.~Dhar and B.~Ratra, {\it Phys. Rev.} {\bf D33}
(1986) 2824\par
\biblitem{PDV} L.~Bonora, M.~Bregola, K.~Lechner, P.~Pasti and M.~Tonin, {\it
Int. J. Mod. Phys.} {\bf A5} (1990) 461\par
\biblitem{ST} A.~Shapiro and C.~C.~Taylor, {\it Phys. Lett.} {\bf 181B} (1986)
67;
 {\bf 186B} (1987) 69\par
\biblitem{PDA} L.~Bonora, P.~Pasti and M.~Tonin, {\it Nucl. Phys.} {\bf B286}
(1987) 150\par
\biblitem{BPT1} L.~Bonora, P.~Pasti and M.~Tonin, {\it Phys. Lett.} {\bf 188B}
(1987) 335\par
\biblitem{PDV1} L.~Bonora, M.~Bregola, K.~Lechner, P.~Pasti and M.~Tonin, {\it
Nucl. Phys.} {\bf B296} (1988) 877\par
\biblitem{TOR3} R.~D'Auria and P.~Fr\'e, {\it Phys. Lett.} {\bf 200B} (1988)
63; R.~D'Auria, P.~Fr\'e, M.~Raciti and F.~Riva {\it Int. J. Mod. Phys.}
{\bf A3} (1988) 953; L.~Castellani, R.~D'Auria and P.~Fr\'e, {\it Phys. Lett.}
{\bf 196B} (1987) 349\par
\biblitem{LPT} K.~Lechner, P.~Pasti and M.~Tonin, {\it Mod. Phys. Lett.}
{\bf A2} (1987) 929\par
\biblitem{LP} K.~Lechner and P.~Pasti, {\it Mod. Phys. Lett.} {\bf A4} (1989)
1721\par
\biblitem{PPT} F.~Paccanoni, P.~Pasti and M.~Tonin, {\it Mod. Phys. Lett.}
{\bf A4} (1989) 807\par
\biblitem{B2AN} M.~B.~Green and H.~H.~Schwarz, {\it Phys. Lett.} {\bf 149B}
(1984) 117\par
\biblitem{B6AN} S.~J.~Gates and H.~Nishino, {\it Phys. Lett.} {\bf 157B}
(1985) 157;
 A.~Salam and E.~Sezgin, {\it Phys. Scr.} {\bf 32} (1985) 283\par
\biblitem{GNISH} S.~J.~Gates and H.~Nishino, {\it Phys. Lett.} {\bf 173B}
(1986) 46\par
\biblitem{WANOM} L.~Bonora, P.~Pasti and M.~Tonin, {\it Phys. Lett.} {\bf 156B}
(1985) 341; {\it Nucl. Phys.} {\bf B261} (1985) 241; {\it Phys. Lett.} {\bf
167B} (1986) 191\par
\biblitem{ETF} L.~Bonora and P.~Cotta--Ramusino, {\it Phys. Lett.} {\bf 107B}
 (1981) 87;
 R.~Stora, {\it Carg\`ese Lectures} (1983);
 B.~Zumino, {\it Les Houches Lectures} (1983);
 B.~Zumino, Y.~S.~Wu and Z.~Zee, {\it Nucl. Phys.} {\bf B239} (1984) 477\par
\biblitem{CL1} A.~Candiello and K.~Lechner, {\it in preparation}\par
\biblitem{CL2} A.~Candiello and K.~Lechner, {\it in preparation}\par
\biblitem{TOR} L.~Castellani, R.~D'Auria and P.~Fr\'e, {\it Supergravity and
Superstrings --- A Geometric Perspective}, World Scientific, Singapore
(1991)\par
\biblitem{DF} R.~D'Auria and P.~Fr\'e, {\it Mod. Phys. Lett.} {\bf A3} (1988)
673\par
\insertbibliografia
\vfill
\eject


\vsize=25truecm
\baselineskip 16truept
\nopagenumbers
\rightline{DFPD/93/TH/51}
\rightline{hep-th/9309143}
\vskip 1truecm
\centerline{{\bf DUALITY IN SUPERGRAVITY THEORIES}\footnote{*}{Supported
in part by M.P.I. This work is carried out in the framework of
the European Community Programme ``Gauge Theories, Applied Supersymmetry
and Quantum Gravity'' with a financial contribution under contract SC1-CT92
-D789.}
}

\vskip 1truecm

\centerline
{\bf Antonio Candiello and Kurt Lechner}

\vskip 1truecm

\centerline
{\sl Dipartimento di Fisica, Universit\`a di Padova}

\smallskip \centerline{\sl and}

\centerline
{\sl Istituto Nazionale di Fisica Nucleare, Sezione di Padova}

\smallskip\centerline{Italy}

\vskip 2truecm

\centerline{\bf Abstract}

\vskip 1truecm
\noindent
The target space dynamics of supermembrane (and superstring) theories
is described by supergravity theories. Supergravity theories associated
to dual supermembrane theories live in the same space-time dimension
and are themselves dual to each other. We present a unified treatment
in superspace of the two dual
formulations of $D=10$, $N=1$ {\it pure}
supergravity based on a strictly
super-geometrical framework: the only fundamental objects are
the
super Riemann curvature and
torsion, and the related Bianchi identities are sufficient to set
the theory on shell; there is no need to introduce, from the beginning,
 closed three- or seven-superforms.
This formulation extends also to {\it non minimal} models.
Moreover, in this framework
the algebraic analogy between pure super Yang--Mills theories and pure
supergravity in $D=10$ is manifest. As an additional outcome in the present
formulation the supersymmetric partner of the ABBJ-Lorentz anomaly in
pure
$D=10$ supergravity can be computed in complete analogy to the ABBJ-gauge
anomaly in
supersymmetric Yang--Mills theories in ten dimensions. In the same
framework  we attack the issue of duality in $N=1$, $D=11$ supergravity
showing in detail that duality holds at the kinematical level in superspace
while it is broken by the dynamics. We discuss also possible extensions of this
theory which could be related to quantum corrections of the eleven dimensional
membrane.
\vfill\eject
\bye